\documentclass{Interspeech}

\interspeechcameraready

\title{Patient-Aware Feature Alignment for Robust Lung Sound Classification: Cohesion-Separation and Global Alignment Losses}

\author[affiliation={1}]{Seung Gyu}{Jeong}
\author[affiliation={1*}]{Seong Eun}{Kim}

\affiliation{}{Seoul National University of Science and Technology}{South Korea}
\email{wa975@naver.com, sekim@seoultech.ac.kr}

\usepackage{comment}
\usepackage{cite}
\usepackage{amsmath,amssymb,amsfonts}
\usepackage{algorithmic}
\usepackage{graphicx}
\usepackage{textcomp}
\usepackage{xcolor}
\usepackage{stfloats}
\usepackage{multirow}
\usepackage{cellspace}
\usepackage{graphicx}
\usepackage{balance}
\usepackage{array}
\newcolumntype{C}[1]{>{\centering\arraybackslash}p{#1}}

\begin{document}

\maketitle

\begin{abstract}
     Lung sound classification is vital for early diagnosis of respiratory diseases. However, biomedical signals often exhibit inter-patient variability even among patients with the same symptoms, requiring a learning approach that considers individual differences. We propose a Patient-Aware Feature Alignment (PAFA) framework with two novel losses, Patient Cohesion-Separation Loss (PCSL) and Global Patient Alignment Loss (GPAL). PCSL clusters features of the same patient while separating those from other patients to capture patient variability, whereas GPAL draws each patient's centroid toward a global center, preventing feature space fragmentation. Our method achieves outstanding results on the ICBHI dataset with a score of 64.84\% for four-class and 72.08\% for two-class classification. These findings highlight PAFA's ability to capture individualized patterns and demonstrate performance gains in distinct patient clusters, offering broader applications for patient-centered healthcare.
\end{abstract}

{
\renewcommand{\thefootnote}{}
\footnote{* Corresponding author. \\ 
Code is available at \url{https://github.com/wa976/PAFA.git}%
} 

\section{Introduction}
Lung sound analysis is essential in early respiratory disease detection, offering a non-invasive means to identify critical conditions such as chronic obstructive pulmonary disease (COPD), asthma, and pneumonia \cite{jacome2015computerized,chambres2018automatic}. However, analyzing these sounds remains challenging due to the diverse patient-specific characteristics that can influence acoustic patterns. Even patients diagnosed with the same condition may exhibit markedly different auscultation signatures arising from anatomical structure variations, pathology severity, and recording environments \cite{shuker2015intra,sapey2008inter}.  

Early approaches to lung sound classification predominantly relied on hand-crafted features and classic machine learning~\cite{sengupta2016lung,ullah2021automatic}, which often struggled to capture the complex spectral and temporal patterns of respiratory signals. More recent efforts have adopted deep convolutional neural networks (CNNs)~\cite{bardou2018lung,petmezas2022automated} or transformer-based models~\cite{bae23b_interspeech,kim2024stethoscope}, yielding improvements in representation learning. However, these models are frequently optimized for class distinctions (e.g., Normal vs. Abnormal), and risk overlooking subtle patient-level distinctions. Some previous studies address inter-patient variability by including demographic data such as age or gender~\cite{moummad2023pretraining,kim24f_interspeech}. However, such metadata alone cannot fully account for the wide range of acoustic differences observed in clinical recordings. 


A further complication arises when global classification objectives (e.g., cross-entropy) dominate model training, potentially causing unique patient embeddings to converge toward larger clusters that do not reflect individual characteristics~\cite{papyan2020prevalence}. This ``Feature Collapse" may yield decent performance on well-represented profiles but neglect outlier cases and their variation. Hence, a key challenge lies in balancing leveraging cross-patient similarities for robust generalization and preserving the distinct identities of each patient's signatures. 

To address these issues, we propose a \textit{Patient-Aware Feature Alignment (PAFA)} framework that explicitly incorporates patient-specific information into the learning process. PAFA aims to preserve individuality while allowing meaningful grouping of shared class-level features. Our approach comprises two complementary cost functions:
\begin{itemize}
    \item \textbf{Patient Cohesion-Separation Loss (PCSL):} This cost promotes cohesion among embeddings from the same patient and enforces separation across different patients, thereby retaining critical patient-specific traits. 
    \item \textbf{Global Patient Alignment Loss (GPAL):} This objective aligns all patient centroids toward a global center, preventing excessive fragmentation in the feature space and maintaining an overarching structure conducive to classification.
\end{itemize}

This combination counters trade-offs in patient-specific modeling. PCSL alone might over-separate patient embeddings, while GPAL alone could underemphasize individual differences by over-mixing different patient representations. The balancing between specificity and generalization prevents either extreme from degrading performance.

We integrate PAFA into the BEATs transformer backbone~\cite{pmlr-v202-chen23ag}, pretrained on AudioSet~\cite{gemmeke2017audio}. We validate our approach using the ICBHI 2017 Challenge Respiratory Sound Database~\cite{rocha2018alpha}. Our experiments show that PAFA consistently outperforms conventional methods in two- and four-class settings while providing particularly pronounced benefits for patients whose acoustic profiles deviate from the majority. These findings highlight the value of explicitly modeling inter-patient variability in biomedical audio applications, paving the way for more effective, personalized diagnostic tools.


\begin{figure*}[t!]
  \centering
  \includegraphics[width=0.9\linewidth]{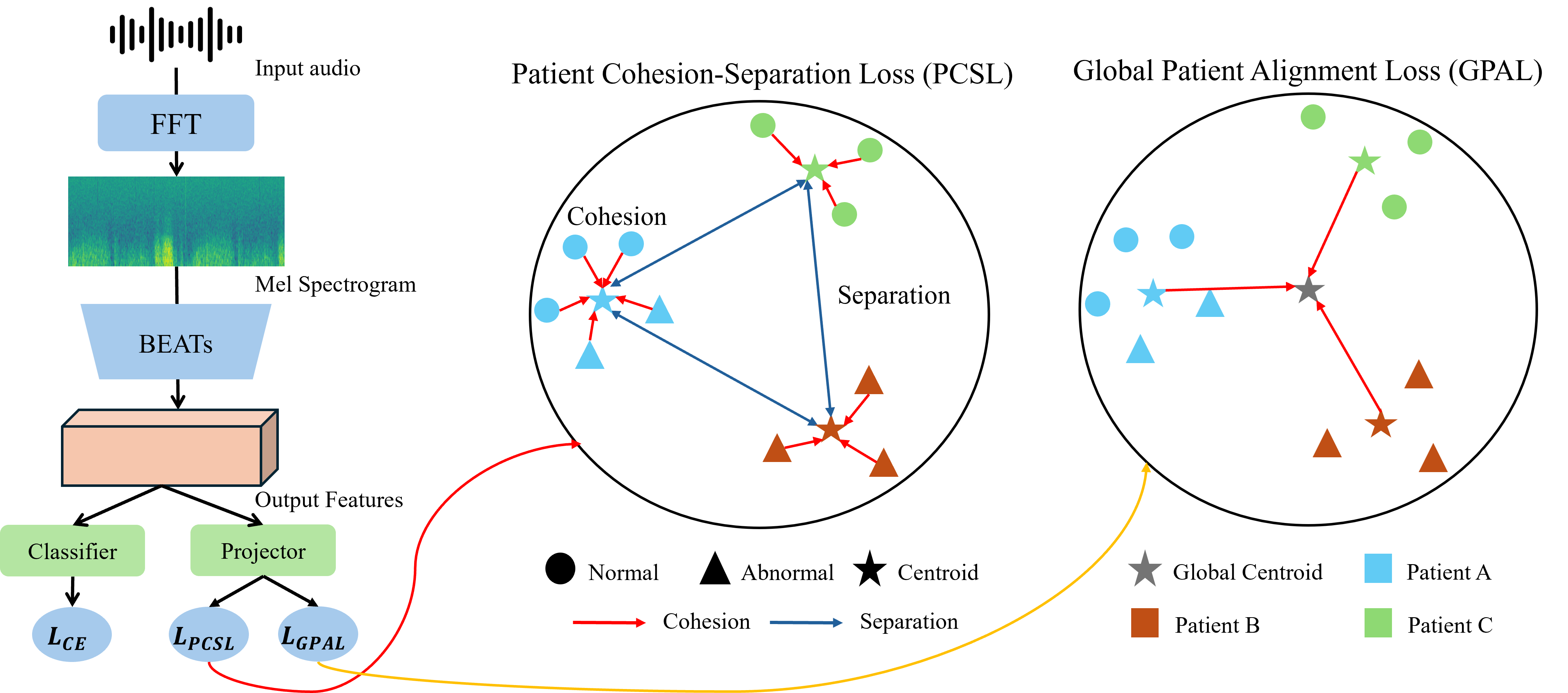}
  \caption{ Overview of the PAFA framework. The input audio is transformed into a Mel Spectrogram. The BEATs backbone encodes the spectrogram into feature embeddings. The projection head computes patient-aware losses (PCSL and GPAL).}
  \label{fig:model}
  \vspace{-0.5cm}
\end{figure*}

\section{Method}

\subsection{PAFA Framework}
Our proposed PAFA framework builds upon the BEATs model as its backbone. A projection head is introduced to impose patient-centric constraints during training through two specialized losses parallel to the primary classification head. Figure~\ref{fig:model} illustrates the complete architecture. By balancing the traditional cross-entropy objective with patient-level constraints, PAFA captures patient-specific acoustic patterns while maintaining clear separation among sound classes.

\subsection{Loss Functions}
The overall training objective comprises the standard cross-entropy loss and two novel patient-aware losses designed to maintain patient-specific consistency while maintaining a coherent global alignment by
\begin{equation}
\mathcal{L}_{\text{total}} = \mathcal{L}_{\text{CE}} + \lambda_{\text{pcsl}}\, \mathcal{L}_{\text{PCSL}} + \lambda_{\text{gpal}}\, \mathcal{L}_{\text{GPAL}}
\label{eq:total_loss}
\end{equation}
where $\mathcal{L}_{\text{CE}}$ is the classification loss, and $\lambda_{\text{pcsl}}$, $\lambda_{\text{gpal}}$ are weighting coefficients for the respective patient-aware losses. Because PCSL and GPAL fulfill complementary roles (patient-level separation vs. global alignment), tuning these weights is crucial. In our study, a grid search determined that $\lambda_{\text{pcsl}}=50$ and $\lambda_{\text{gpal}}=0.0005$ yield the best performance.

\subsubsection{Patient Cohesion-Separation Loss (PCSL)}
PCSL promotes intra-patient consistency by clustering feature representations from the same patient while enforcing sufficient separation between different patients. $\mathbf{z}_i$ is feature embedding for the $i$-th sample. For each patient $p \in \mathcal{P}$, where $\mathcal{P}$ is the set of all patients in the current batch, centroid $\mu_p$ measured by
\begin{equation}
\mu_p = \frac{1}{N_p} \sum_{i \in p} \mathbf{z}_i,
\label{eq:centroid}
\end{equation}
where $N_p$ is the number of samples for patient $p$. Intra-patient compactness $S_W$ is quantified by
\begin{equation}
S_W = \sum_{p \in \mathcal{P}} \sum_{i \in p} \left\lVert \mathbf{z}_i - \mu_p \right\rVert_2^2.
\label{eq:sw}
\end{equation}
A lower value of $S_W$ indicates that samples from the same patient are tightly clustered. In contrast, inter-patient dispersion $S_B$ is measured by
\begin{equation}
S_B = \sum_{\substack{p,q \in \mathcal{P} \\ p \neq q}} \left\lVert \mu_p - \mu_q \right\rVert_2^2.
\label{eq:sb}
\end{equation}
A larger $S_B$ implies well-separated patient centroids, thus aiding in distinguishing among different patients.

PCSL is conceptually inspired by Fisher’s Linear Discriminant Analysis (LDA) \cite{li2014fisher}, which seeks to maximize between-class variance while minimizing within-class variance. Here, we replace the notion of “class” with “patient” to prevent the model from collapsing all patients’ features into a dominant cluster. Concretely, $\mathcal{L}_{\text{PCSL}}$ is formulated as
\begin{equation}
\mathcal{L}_{\text{PCSL}} = \frac{S_W}{S_B + \epsilon},
\label{eq:pcsl}
\end{equation}
where $\epsilon$ is a small constant for numerical stability. Minimizing this ratio draws samples of the same patient closer together while pushing different patient centroids apart. PCSL helps retain distinct embedding regions for less-represented patients, preventing from merging with those of larger patient groups.

\subsubsection{Global Patient Alignment Loss (GPAL)}
While PCSL encourages clear patient-level clustering, it can cause these clusters to drift excessively within the embedding space, potentially compromising the overall class structure. To counteract this, GPAL aligns all patient centroids toward a common global reference. We compute the global centroid $\mu_G$ as the mean of all patient centroids $\mu_p$ in the batch by
\begin{equation}
\mu_G = \frac{1}{|\mathcal{P}|} \sum_{p \in \mathcal{P}} \mu_p,
\label{eq:global_centroid}
\end{equation}
and $\mathcal{L}_{\text{GPAL}}$ is defined by
\begin{equation}
\mathcal{L}_{\text{GPAL}} = \frac{1}{|\mathcal{P}|} \sum_{p \in \mathcal{P}} \left\lVert \mu_p - \mu_G \right\rVert_2^2.
\label{eq:gpal}
\end{equation}
By constraining patient centroids around $\mu_G$, GPAL prevents over-separation while preserving the distinctiveness enforced by PCSL. This global alignment is particularly beneficial when encountering unseen new patients, as it provides a coherent reference in the embedding space for all patient clusters.

\subsection{Training Procedure}
We train the model end-to-end using the combined loss in Equation~\ref{eq:total_loss}. At each iteration, lung sound features are extracted by the BEATs backbone and passed through the projection head to compute PCSL and GPAL. The total loss is backpropagated to update all parameters.

In practice, we group embeddings by patient identifiers to compute the intra- and inter-patient statistics for PCSL, and measure each patient centroid’s distance to the global centroid for GPAL. Once training concludes, the auxiliary projection head for the patient-aware losses is removed, and only the classification head is used during inference.

\begin{table*}[t!]
\centering
\caption{Performance comparison of different methods on the ICBHI dataset. IN, AS, and LA refer to ImageNet \cite{deng2009imagenet}, AudioSet \cite{gemmeke2017audio}, and LAION-Audio-630K \cite{wu2023large}, respectively. All results are averaged values over five runs with different random seeds.}
\label{tab:performance_comparison}
\renewcommand{\arraystretch}{1.15}
\begin{tabular}{p{0.3cm} p{4.3cm} p{1.5cm} >{\centering\arraybackslash}p{1.3cm} >{\centering\arraybackslash}p{1.3cm} | >{\centering\arraybackslash}p{1.2cm} >{\centering\arraybackslash}p{1.2cm} >{\centering\arraybackslash}p{1.4cm}}
\hline
\hline
& Method & Backbone & Pretraining & Metadata & Sp (\%) & Se (\%) & Score (\%) \\
\hline
\hline
\multirow{7}{*}{\rotatebox{90}{4-class}}
    & Gairola et al. (RespireNet) \cite{gairola2021respirenet} & ResNet34 & IN &  X & 72.30 & 40.10 & 56.20 \\
    &Moummad et al. (SCL) \cite{moummad2023pretraining} & CNN6 & AS & O & 75.95 & 39.15 & 57.55 \\
    & Bae et al. (Patch-Mix CL) \cite{bae23b_interspeech} & AST & IN + AS & X & 81.66 & 43.07 & 62.37 \\
    & Kim et al. (SG-SCL) \cite{kim2024stethoscope} & AST & IN + AS & O & 79.87 & 43.55 & 61.71 \\
    &Kim et al. (BTS) \cite{kim24f_interspeech} & CLAP & LA & O & 81.40 & 45.67 & 63.54 \\
    \cline{2-8}
    &\textbf{BEATs + CE [ours]} & BEATs & AS & X & $78.77$\text{\scriptsize$\pm$3.07} & $\textbf{48.21}$\text{\scriptsize$\pm$2.32} & $63.49$\text{\scriptsize$\pm$1.08} \\
    &\textbf{BEATs + PAFA [ours]} & BEATs & AS & X & $\textbf{82.05}$\text{\scriptsize$\pm$1.95}& $47.63$\text{\scriptsize$\pm$2.23}& $\textbf{64.84}$\text{\scriptsize$\pm$0.60}\\
\hline
\multirow{5}{*}{\rotatebox{90}{2-class}} 
    & Nguyen et al. (CoTuning) \cite{nguyen2022lung} & ResNet50 & IN & X & 79.34 & 50.14 & 64.74 \\
    &  Bae et al. (Patch-Mix CL) \cite{bae23b_interspeech} & AST & IN + AS & X & \textbf{81.66} & 55.77 & 68.71 \\
    &  Kim et al. (SG-SCL) \cite{kim2024stethoscope} & AST & IN + AS & O & 79.87 & 57.97 & 68.93  \\
    \cline{2-8}
    &\textbf{BEATs + CE [ours]} & BEATs & AS & X & $75.19$\text{\scriptsize$\pm$6.49} & $66.34$\text{\scriptsize$\pm$5.90} & $70.76$\text{\scriptsize$\pm$0.42} \\
    &\textbf{BEATs + PAFA [ours]} & BEATs & AS & X & $74.87$\text{\scriptsize$\pm$2.64} & $\textbf{68.29}$\text{\scriptsize$\pm$2.78} & $\textbf{72.08}$\text{\scriptsize$\pm$0.55} \\
\hline
\end{tabular}
\end{table*}

\section{Experimental Setup}

\subsection{Dataset}
This study uses the ICBHI 2017 dataset \cite{rocha2018alpha}, comprising 6,898 recordings from 126 subjects. The data is categorized into four classes---Normal, Crackle, Wheeze, and Both---with sampling rates between 4 kHz and 44.1 kHz. The training set (4,142 samples) comprises 2,063 Normal, 1,215 Crackle, 501 Wheeze, and 363 Both recordings, while the test set (2,756 samples) has 1,579 Normal, 649 Crackle, 385 Wheeze, and 143 Both recordings. This dataset presents substantial patient-specific acoustic variability, making cross-subject generalization challenging.

\subsection{Preprocessing}
We followed the preprocessing pipeline outlined for the pretrained BEATs model \cite{pmlr-v202-chen23ag}, with minor modifications for the ICBHI dataset. All recordings were resampled to 16 kHz, segmented into 5-second intervals (shorter recording padded by repetition, longer recordings truncated), and converted to 128-bin fbank features. 

\subsection{Training Details}
The BEATs model was initialized with pretrained weights from the AudioSet dataset \cite{gemmeke2017audio}. Using the Adam optimizer with an initial learning rate of \(5 \times 10^{-5}\) and a weight decay of \(1 \times 10^{-6}\), we trained for 100 epochs with a batch size of 32.

\subsection{Evaluation Metrics}
 We evaluated performance using Sensitivity (Se), Specificity (Sp), and the ICBHI Score (Score) \cite{rocha2018alpha}, defined as $(\text{Se}+\text{Sp})/2$. Sensitivity assesses the detection of respiratory anomalies, and Specificity measures accuracy on normal sounds. Both four-class (Normal, Crackle, Wheeze, Both) and two-class (Normal, Abnormal) setups were tested.

\section{Result}

\subsection{Performance on the ICBHI Dataset}
We evaluated our method on the ICBHI dataset using the official 60/40 (train/test) split. Table~\ref{tab:performance_comparison} shows Sp, Se, and the overall Score for both four- and two-class tasks. Compared with recent approaches, our BEATs-based models achieve state-of-the-art Scores in both tasks. Specifically, for the four-class task, BEATs+PAFA attains a Score of 64.84\%, outperforming our CE-only baseline by 1.35\%. For the two-class task, BEATs+PAFA reaches 72.08\%, exceeding the CE-only baseline by 1.32\%. These results confirm that integrating patient-aware constraints effectively addresses inter-patient variability while preserving unique acoustic traits.

\begin{table}[t]
\centering
\caption{Performance Metrics for Different Backbones.}
\label{tab:performance_metrics}
\resizebox{0.45\textwidth}{!}{%
\begin{tabular}{lllccl}
\toprule
\textbf{Model} & \textbf{Method}  & \textbf{Pretrain} & \textbf{Sp (\%)} & \textbf{Se (\%)} & \textbf{Score (\%)} \\
\midrule
\multirow{2}{*}{CNN6} 
 & CE   & \multirow{2}{*}{IN}      & $\textbf{83.05}$\text{\scriptsize$\pm$2.80} & $31.88$\text{\scriptsize$\pm$3.44} & $57.46$\text{\scriptsize$\pm$0.85} \\
 & PAFA &                          & $82.47$\text{\scriptsize$\pm$3.38} & $\textbf{33.27}$\text{\scriptsize$\pm$4.24} & $\textbf{57.87}$\text{\scriptsize$\pm$0.75} \\
\midrule
\multirow{2}{*}{AST} 
 & CE   & \multirow{2}{*}{IN + AS}   & $79.18$\text{\scriptsize$\pm$5.43} & $42.68$\text{\scriptsize$\pm$5.04} & $60.93$\text{\scriptsize$\pm$0.50} \\
 & PAFA &                           & $\textbf{79.29}$\text{\scriptsize$\pm$1.79} & $\textbf{45.08}$\text{\scriptsize$\pm$1.78} & $\textbf{62.19}$\text{\scriptsize$\pm$0.89} \\
\midrule
\multirow{2}{*}{BEATs} 
 & CE   & \multirow{2}{*}{AS}        & $78.77$\text{\scriptsize$\pm$3.07} & $\textbf{48.21}$\text{\scriptsize$\pm$2.32} & $63.49$\text{\scriptsize$\pm$1.08} \\
 & PAFA &                           & $\textbf{82.05}$\text{\scriptsize$\pm$1.95}& $47.63$\text{\scriptsize$\pm$2.23}& $\textbf{64.84}$\text{\scriptsize$\pm$0.60}\\
\bottomrule
\end{tabular}%
}
\end{table}

\begin{table}[t]
\centering
\caption{Ablation study on the effect of PCSL and GPAL}
\label{tab:ablation_study}
\renewcommand{\arraystretch}{1.2} 
\resizebox{0.45\textwidth}{!}{%
\begin{tabular}{lccc} 
\toprule 
Variant & Se(\%) & Sp(\%) & Score(\%) \\
\midrule 
Full method & $\textbf{82.05}$\text{\scriptsize$\pm$1.95}& $47.63$\text{\scriptsize$\pm$2.23}& $\textbf{64.84}$\text{\scriptsize$\pm$0.60}\\
w/o PCSL & $76.29$\text{\scriptsize$\pm$2.91} & $51.56$\text{\scriptsize$\pm$1.60} &$63.92$\text{\scriptsize$\pm$0.92}  \\
w/o GPAL & $75.83$\text{\scriptsize$\pm$3.54} & $\textbf{51.93}$\text{\scriptsize$\pm$2.75} &$63.88$\text{\scriptsize$\pm$0.72}  \\
\bottomrule 
\end{tabular}%
}
\end{table}

\subsection{Backbone Comparison}
To ensure that PAFA's benefits are not tied to a specific model, we tested three different backbones: CNN6, AST, and BEATs \cite{kong2020panns,gong21b_interspeech,pmlr-v202-chen23ag}. As shown in Table~\ref{tab:performance_metrics}, PAFA increases the Score by 0.41\% for CNN6 (from 57.46\% to 57.87\%), by 1.26\% for AST (60.93\% to 62.19\%), and by 1.35\% for BEATs (63.49\% to 64.84\%). These consistent improvements confirm the broad applicability of patient-centric losses.

\begin{figure}[t!]
    \centering
    \includegraphics[width=0.47\linewidth]{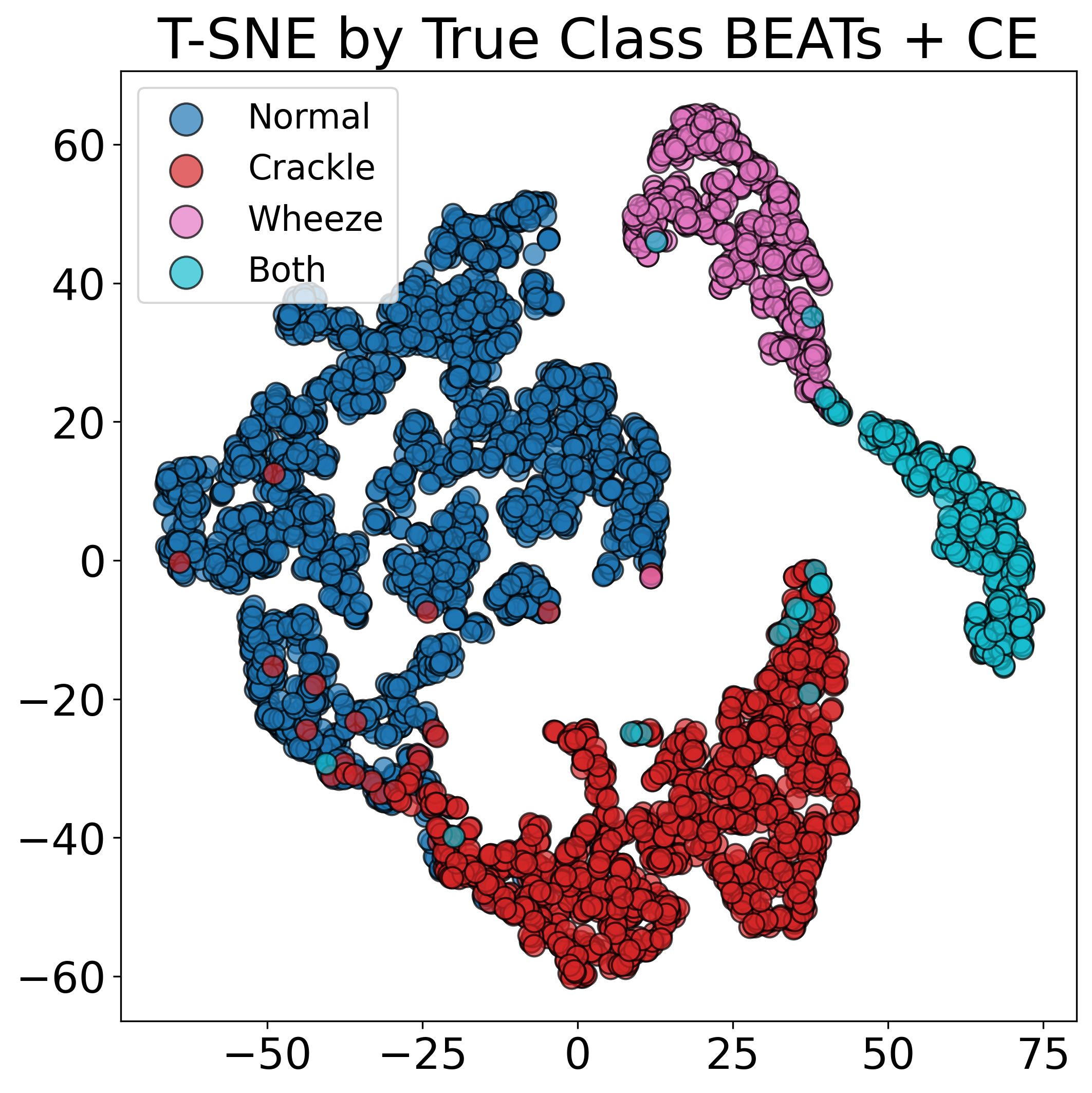}
    \hspace{3mm}
    \includegraphics[width=0.47\linewidth]{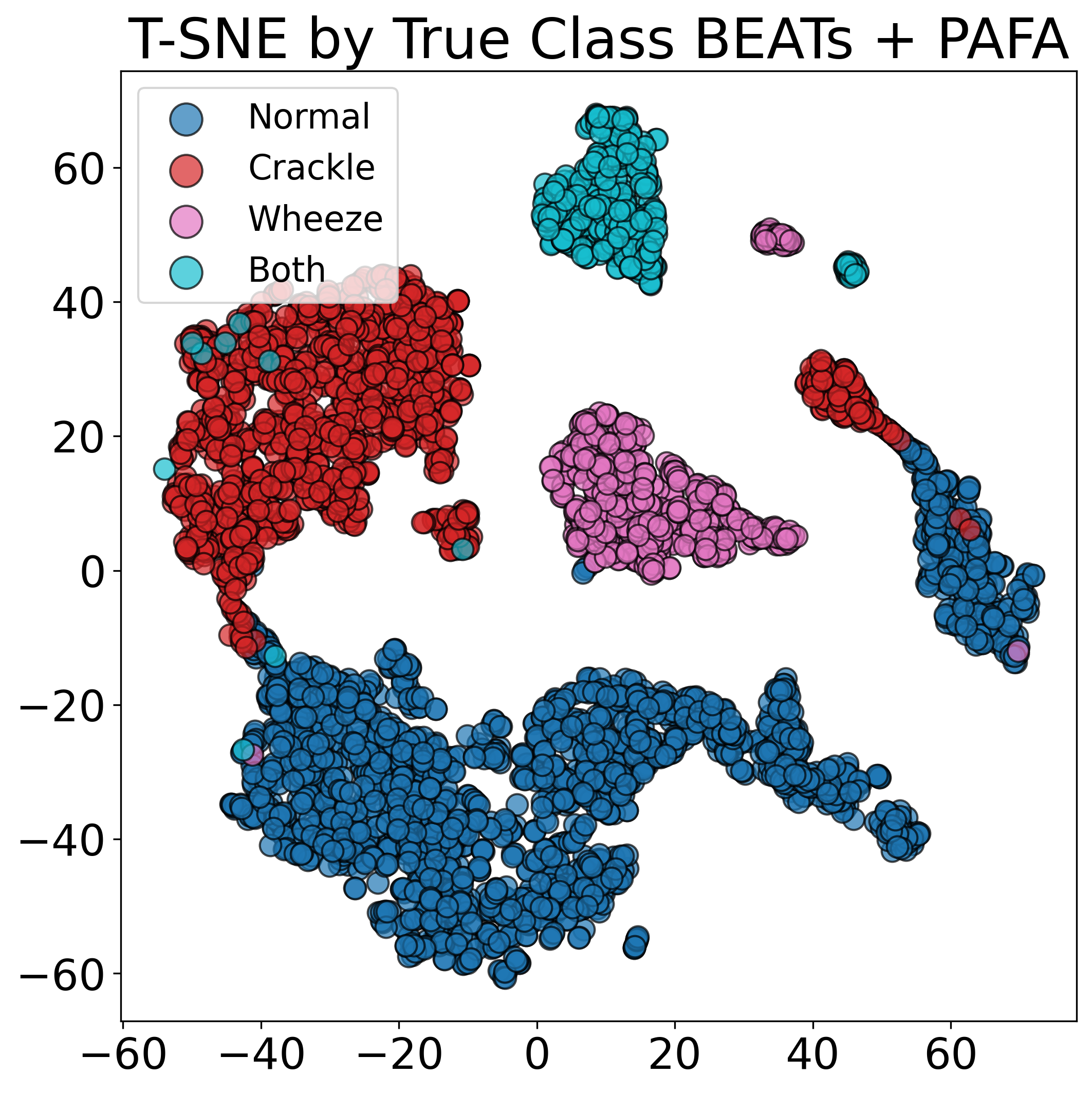}
    \caption{t-SNE visualization of feature embeddings from the training data. (Left) CE-only. (Right) PAFA.}
    \label{fig:tsne_class_comparison}
\end{figure}

\begin{figure}[t!]
    \centering
    \includegraphics[width=0.47\linewidth]{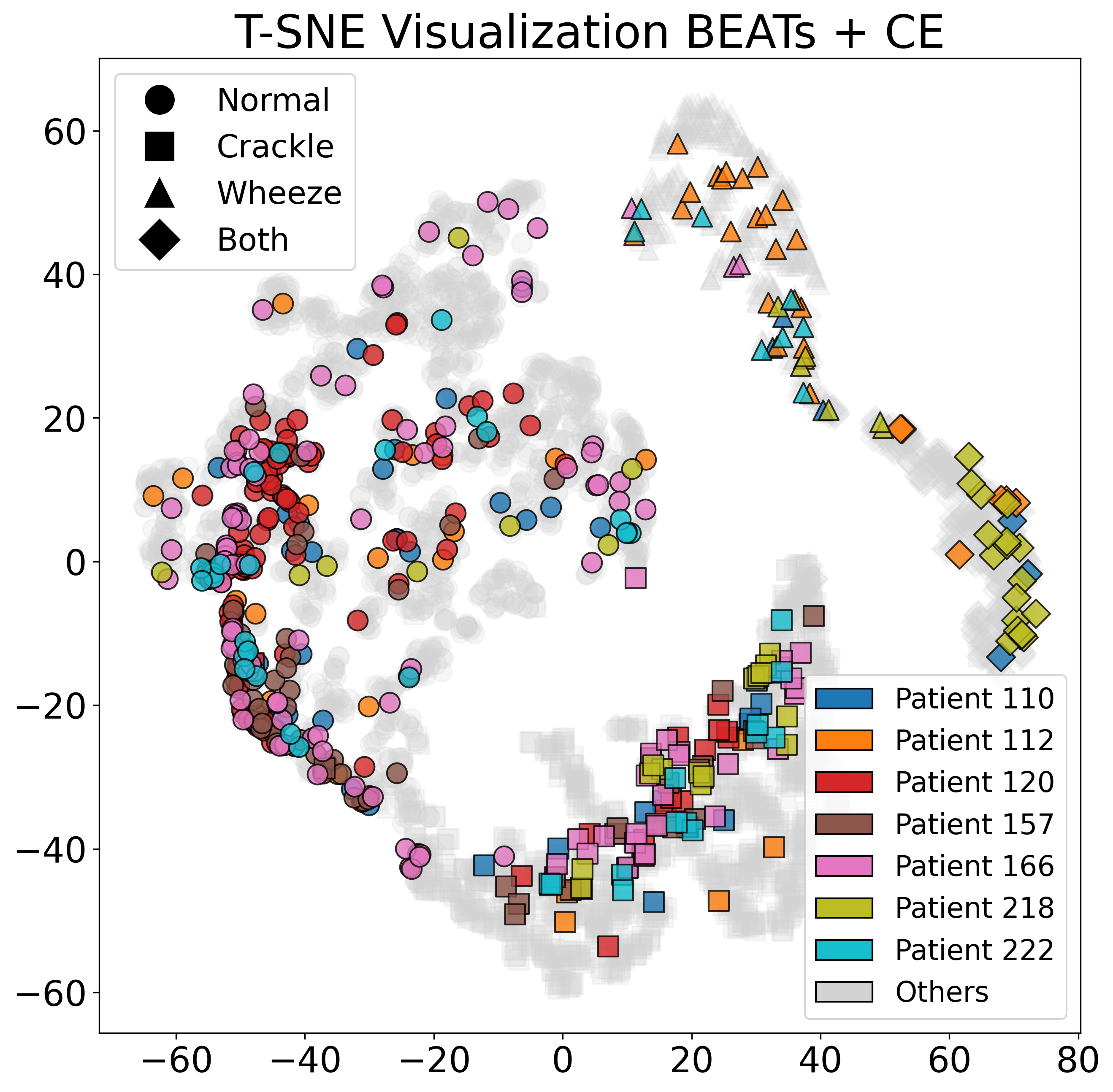}
    \hspace{3mm}
    \includegraphics[width=0.47\linewidth]{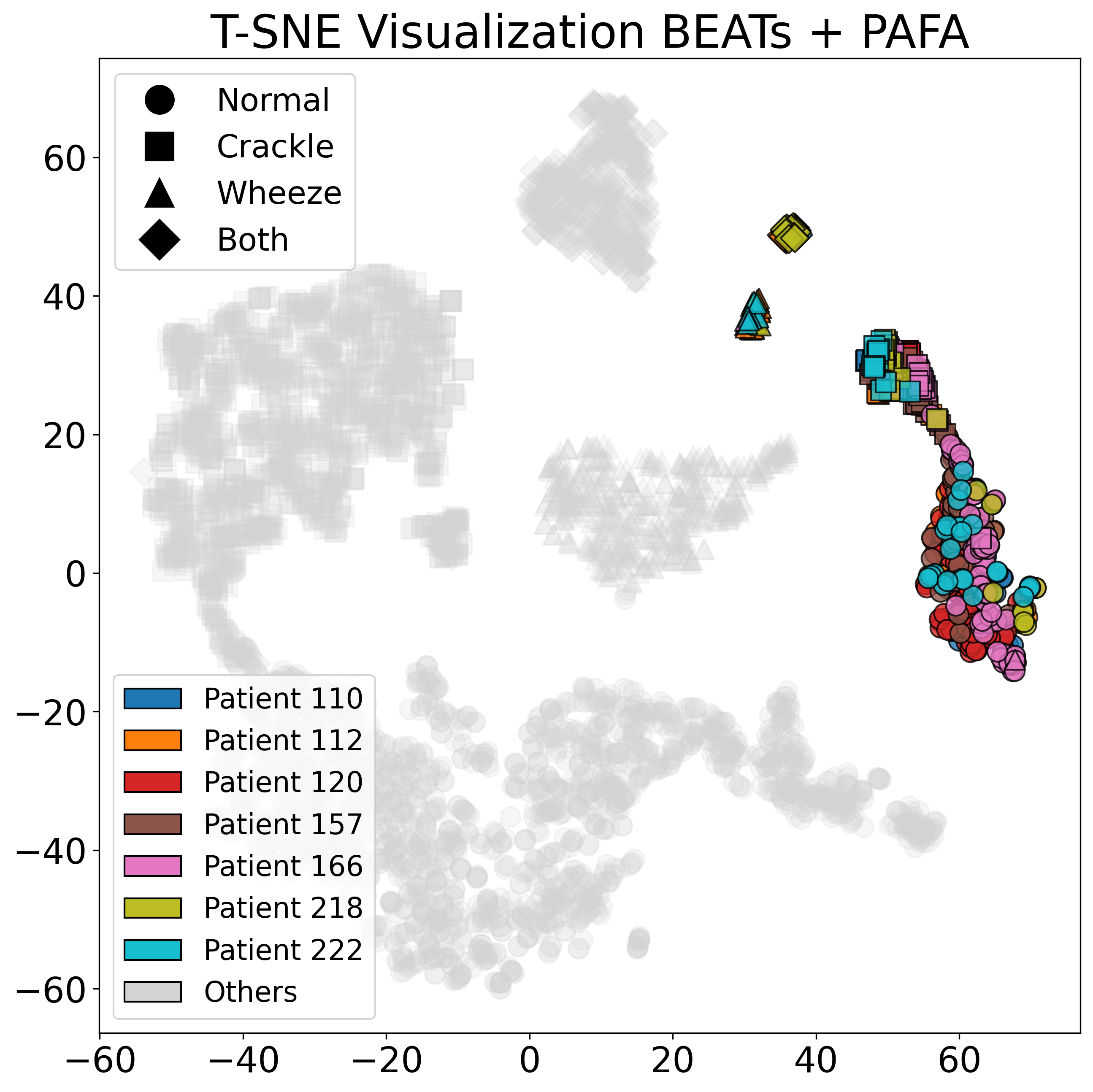}
    \caption{t-SNE visualization focusing on training samples from a subset of patients who form new sub-clusters under PAFA. (Left) CE-only. (Right) PAFA.}
    \label{fig:tsne_subclusters}
\end{figure}

\subsection{Ablation Study on Patient-Aware Loss Components}
We conducted an ablation study to evaluate the separate contributions of PCSL and GPAL. Table~\ref{tab:ablation_study} shows that removing either component reduces the Score to approximately 63.90\%, whereas the entire PAFA framework achieves 64.84\%. Although each partial configuration still outperforms the CE-only baseline, the complete approach yields the best overall performance. This result indicates that \textit{PCSL} and \textit{GPAL} are complementary. PCSL promotes intra-patient cohesion, and GPAL enforces global alignment among patient clusters.

\subsection{The Impact of PAFA on Test Data}

Figure~\ref{fig:tsne_class_comparison} compares t-SNE embeddings of the training data under both CE-only and PAFA. With CE-only, most samples from four large clusters aligned with the primary respiratory sound classes. In contrast, PAFA reveals additional sub-clusters, indicating that certain patient groups are no longer subsumed within a single class cluster. In Figure~\ref{fig:tsne_subclusters}, we focus on a subset of patients forming distinct sub-clusters under PAFA. With CE-only, these patients typically merge into the major class-centric clusters. However, under PAFA, they occupy more individualized regions in the embedding space, avoiding collapse into dominant distributions.

Figure~\ref{fig:tsne_high} presents another perspective by examining patients with the largest number of samples. In both CE-only and PAFA, these highly sampled patients still tend to reside within the main class-centric clusters, presumably because their abundant data ensures robust representation in either method.

Figure~\ref{fig:new_barplots} shows two side-by-side bar plots comparing CE vs PAFA for selected test patients. In the left bar plot, we first compute the centroid of each newly formed sub-cluster's patients (from Figure~\ref{fig:tsne_subclusters}) and select six test patients whose feature representations are closest to these centroids. Although some patients exhibit minor performance drops under PAFA, others improve by a large margin. In the right bar plot, we use the high-sample patients (from Figure~\ref{fig:tsne_high}) and select the six test patients whose features are closest to them. Here, we observe only a slight increase or decrease, indicating that overall performance remains relatively stable. Overall, by allowing patient-specific clustering in the feature space, PAFA effectively addresses variations in patient profiles during the test phase, significantly improving performance.

\begin{figure}[t!]
    \centering
    \includegraphics[width=0.47\linewidth]{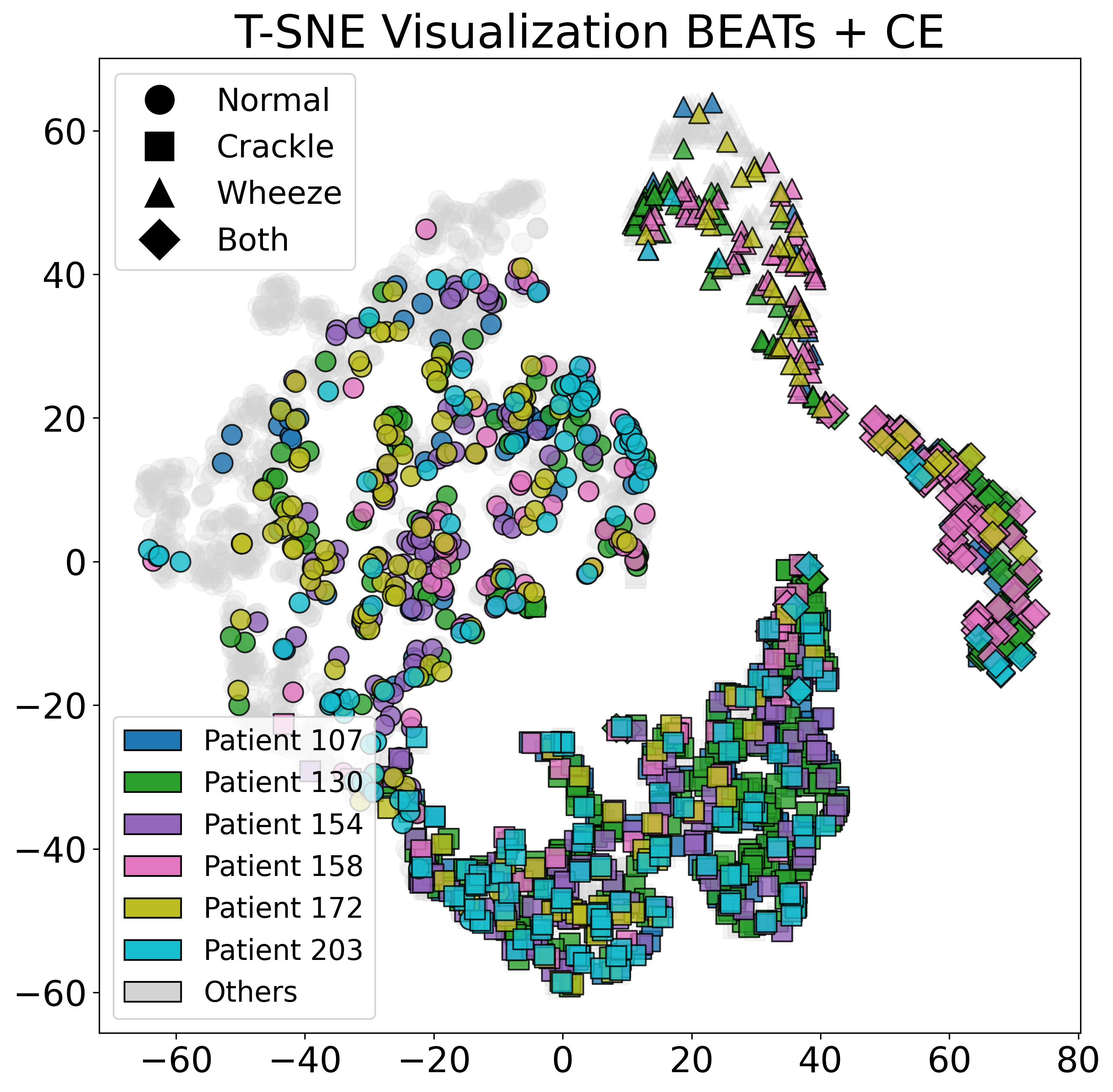}
    \hspace{3mm}
    \includegraphics[width=0.47\linewidth]{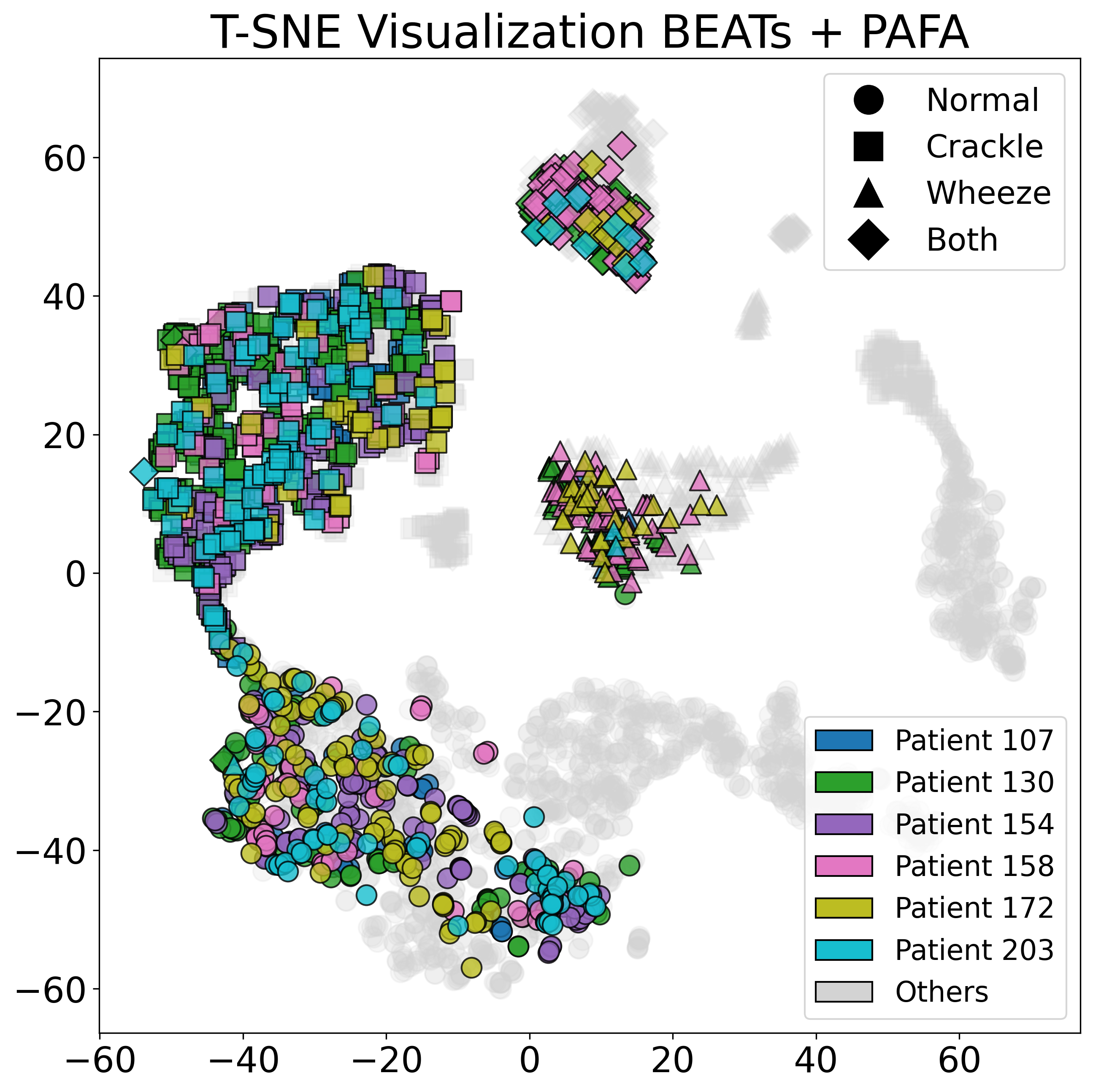}
    \caption{t-SNE visualization of training samples from patients with the highest number of recordings. (Left) CE-only. (Right) PAFA.}
    \label{fig:tsne_high}
\end{figure}

\begin{figure}[t!]
    \centering
    \includegraphics[width=0.47\linewidth]{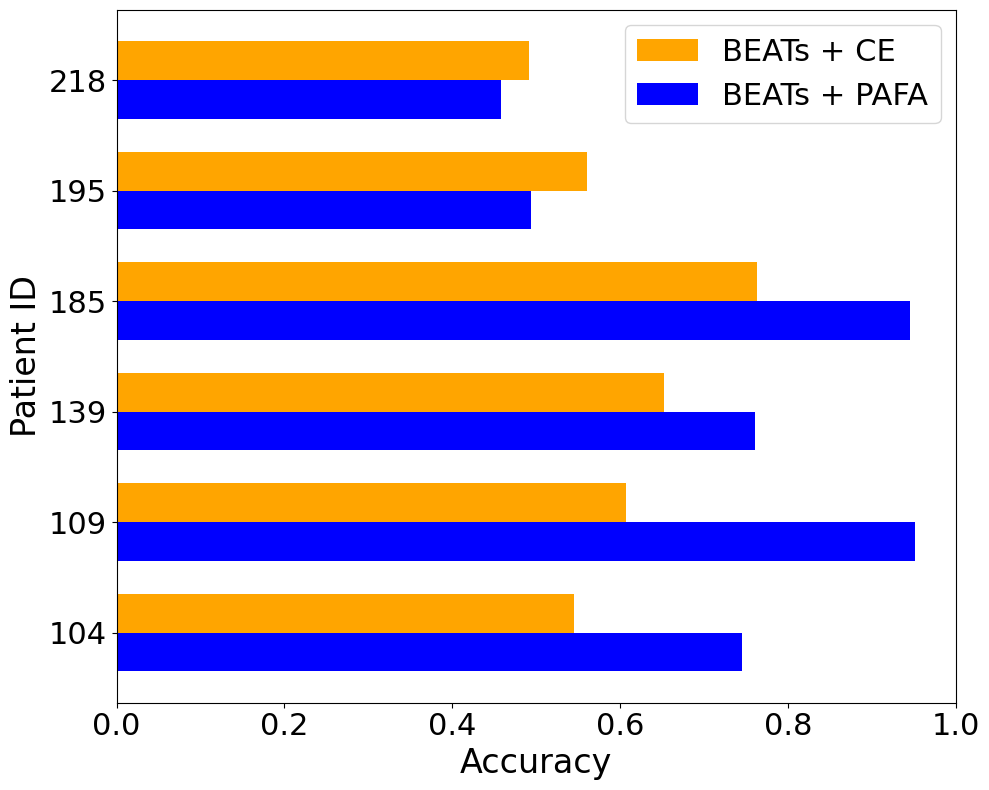}
    \hspace{3mm}
    \includegraphics[width=0.47\linewidth]{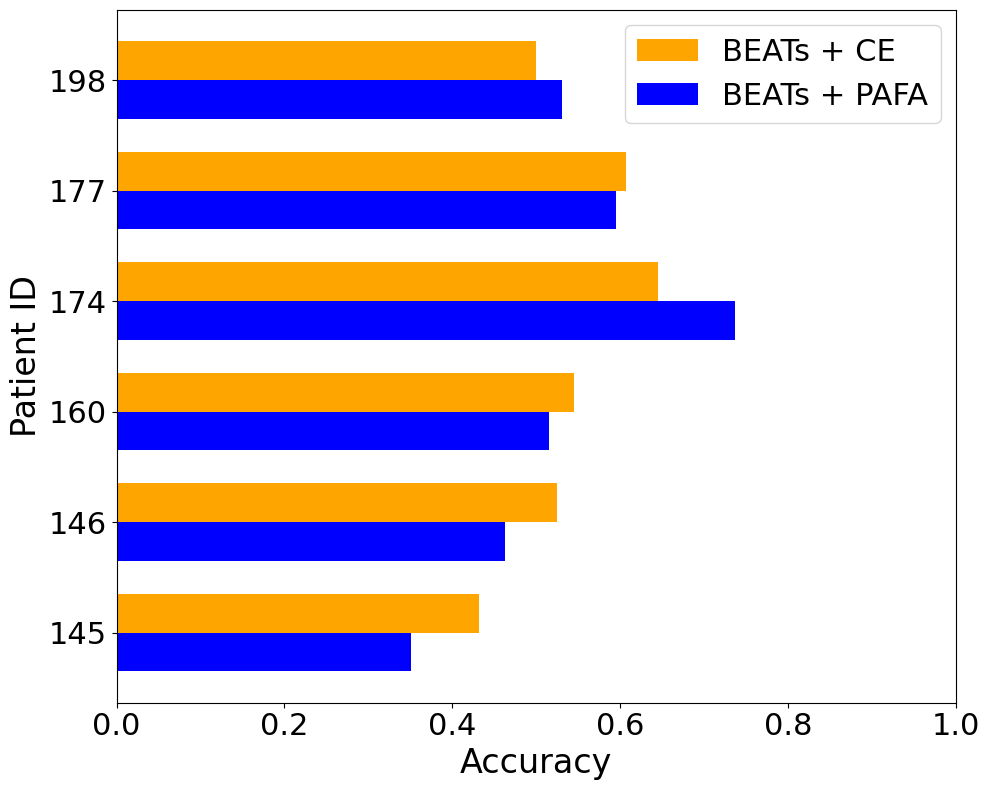}
    \caption{Patient accuracy comparison (CE vs.\ PAFA) on selected test patients. (Left) Six test patients nearest to centroids of new sub-clusters (from PAFA training embeddings). (Right) Six test patients nearest to centroids of high-sample patients (from training embeddings).}
    \label{fig:new_barplots}
\end{figure}

\section{Conclusion}

We introduced Patient-Aware Feature Alignment (PAFA), a framework that explicitly incorporates patient-centric constraints into lung sound classification. By simultaneously applying Patient Cohesion-Separation Loss (PCSL) and Global Patient Alignment Loss (GPAL), PAFA enforces both intra-patient cohesion and an overarching global structure, ensuring that subtle patient-specific traits remain distinct rather than merging into larger, dominant clusters. Evaluations on the ICBHI dataset demonstrated that our method improves classification performance across various patient profiles, confirming that PAFA effectively mitigates the collapse of underrepresented patient groups. By preserving these subtle variations, the model becomes better equipped to handle novel test patients. In future work, we plan to extend PAFA for dynamic adaptation to unseen patient data, facilitating real-time updates in continuously evolving clinical environments.

\section{Acknowledgement}
This work was supported in part by the National Research Foundation of Korea (NRF) funded by the Ministry of Science and ICT (MSIT) of the Korean government under Grant RS-2023-00221365 and in part by Seoul National University of Science and Technology.

\bibliographystyle{IEEEtran}
\bibliography{mybib}

\end{document}